\documentclass[twocolumn]{aastex631}
\usepackage{amsmath}
\usepackage{amssymb}
\usepackage{physics}
\usepackage{dsfont}
\usepackage{mathrsfs}
\usepackage{wasysym}
\usepackage{multirow}
\usepackage{comment}
\usepackage[outline]{contour}
\usepackage{orcidlink}
\usepackage{hyperref}

\newcommand\Htwo{$\rm H_2$}
\newcommand\Sone{$\rm H_2\,S(1)$}
\newcommand\Sfive{$\rm H_2\,S(5)$}
\newcommand\Snine{$\rm H_2\,S(9)$}

\newcommand\COthree{$\rm CO(3-2)$}
\newcommand\COone{$\rm CO(1-0)$}

\newcommand\Msun{$\rm M_{\odot}$}

\begin{document}

\title{Warm and cold molecular gas in the cluster center of MACS 1931-26 with JWST and ALMA}

\correspondingauthor{Laya Ghodsi}
\email{Layaghodsi@phas.ubc.ca}

\author{L. Ghodsi\,\orcidlink{0009-0008-8936-1625}}
\affiliation{ Department of Physics \& Astronomy, University of British Columbia, 6224 Agricultural Road, Vancouver BC, V6T 1Z1, Canada}

\author{L. Kuhn\,\orcidlink{0009-0005-0766-5026}}
\affiliation{ Department of Physics \& Astronomy, University of British Columbia, 6224 Agricultural Road, Vancouver BC, V6T 1Z1, Canada}

\author{A. W. S. Man\,\orcidlink{0000-0003-2475-124X}}
\affiliation{ Department of Physics \& Astronomy, University of British Columbia, 6224 Agricultural Road, Vancouver BC, V6T 1Z1, Canada}

\author{P. Andreani\,\orcidlink{0000-0001-9493-0169}}
\affiliation{ESO, Karl Schwarzschild strasse 2, 85748 Garching, Germany}

\author{C. De Breuck\,\orcidlink{0000-0002-6637-3315}}
\affiliation{ESO, Karl Schwarzschild strasse 2, 85748 Garching, Germany}

\author{A. Togi\,\orcidlink{0000-0001-5042-3421
}}
\affiliation{Department of Physics, Texas State University, 601 University Drive, San Marcos, TX 78666, USA}

\author{K. Dasyra\,\orcidlink{0000-0002-1482-2203
}}
\affiliation{Section of Astrophysics, Astronomy \& Mechanics, Department of Physics,  National and Kapodistrian University of Athens, Panepistimioupolis Zografou, 15784 Athens, Greece}

\author{M. Lehnert\,\orcidlink{0000-0003-1939-5885}}
\affiliation{ Centre de Recherche Astrophysique de Lyon, ENS de Lyon, Université Lyon 1, CNRS, UMR5574, 69230 Saint-Genis-Laval,
France}

\author{I. García-Bernete\,\orcidlink{0000-0002-9627-5281}}
\affiliation{Department of Physics, University of Oxford, Oxford OX1 3RH, UK}

\author{D. Donevski\,\orcidlink{0000-0001-5341-2162}}
\affiliation{National Centre for Nuclear Research, Pasteura 7, 02-093 Warsaw, Poland}
\affiliation{SISSA, Via Bonomea 265, 34136 Trieste, Italy}

\author{T.~G. Bisbas\,\orcidlink{0000-0003-2733-4580}}
\affiliation{Research Center for Astronomical Computing, Zhejiang Laboratory, Hangzhou 311100, China}

\author{Y. Miyamoto\,\orcidlink{0000-0002-7616-7427}}
\affiliation{Fukui University of Technology, 3-6-1, Gakuen, Fukui City, Fukui Pref. 910-8505, Japan}

\begin{abstract}

We perform one of the first spatially resolved studies of warm ($>$100 K) and cold (10–100 K) molecular gas in the circumgalactic medium (CGM), focusing on the brightest cluster galaxy (BCG) of a cool-core galaxy cluster, MACS1931-26 at z=0.35. This galaxy has a massive \Htwo~reservoir and a radio-loud active galactic nucleus (AGN) and is undergoing a starburst event. We present new JWST observations of this system, revealing warm \Htwo~gas that is co-spatial with the cold molecular gas traced by CO, extending over 30 kpc around the BCG in a tail-like structure reaching into the circumgalactic medium of this galaxy. Analysis of the mid-infrared pure \Htwo~rotational lines \Sone, \Sfive, and \Snine~indicate warm gas temperatures of $515.6 \pm 0.8$ K and $535.2\pm 1.9$ K in the BCG and tail regions, respectively. We compare cold gas, traced by the \COthree~observed with ALMA, to the warm gas traced by JWST. The warm-to-cold molecular gas fraction is comparable in the BCG ($1.4\%\pm0.2\%$) and the CGM tail ($1.9\%\pm0.3\%$). Our analysis suggests that the dissipation of the kinetic energy of the \Htwo~emitting gas is sufficient to lead to the formation of the CO gas. This observation provides new insights into the molecular gas distribution and its potential role in the interplay between the central galaxy and its circumgalactic environment.

\end{abstract}


%%%%%%%%%%%%%%%%%%%%%%%%%%%%%%%%%%%%%%%%%%%%%%%%%%%%%%%%%%%%%%%%%%%%%%%%%%%%%%%%%%%%%%%%%%%%%%%%%%%%%%%%%
\section{Introduction}

\label{sec:intro}
Galaxies interact with their surrounding medium across various physical scales. Within the virial radius of galaxies, the circumgalactic medium (CGM) serves as the environment where the exchange of low-metallicity pristine gas from the cosmological environment (the intergalactic medium, IGM, and intracluster medium, ICM for cluster galaxies) and metal-enriched recycled gas from galaxies occurs.  
The CGM acts as a bridge between galaxies and their broader cosmological environment, facilitating gas exchange that influences star formation by providing fresh fuel and redistributing metals \citep{Werk14, McDonald16, Tumlinson17, Voit21}. 

The CGM is a multiphase medium consisting of gas with temperatures ranging from 10 to $10^6$ K, along with dust present as a cooler component. Its thermodynamic and ionization states are shaped by the interplay between gas cooling, heating mechanisms, and feedback from the host galaxy. Understanding the links between the CGM, its host galaxy, and the broader cosmological environment is crucial to unraveling the galactic baryon cycle and the regulation of star formation \citep{Voit15, Faucher23}.

Although CO and [CI] are commonly used tracers of cold molecular hydrogen (10–100 K), a fraction of the gas is warm and cannot be traced by the traditionally-used low-J CO lines \citep{Field66, Wolfire10, Togi16, Seifried20}. Warm gas ($>$100 K) can be detected through the pure rotational transitions of \Htwo~ in the mid-infrared. Possible excitation sources for this phase of molecular gas include shocks from galactic outflows or cloud collisions, far-ultraviolet radiation from massive stars in photodissociation regions, X-ray emission produced by active galactic nuclei (AGN) and supernova remnants, and cosmic rays originating from AGN and hot ICM \citep{Roussel07,Ogle10,  Guillard12, Dasyra22}. In this work, we examine the cold and warm molecular gas flows around the star-forming brightest cluster galaxy (BCG) to better understand the heating and excitation mechanisms governing the dynamics and energetics of these gas flows.

MACS 1931.8-2635 (hereafter MACS1931) is a cool core galaxy cluster at $z=0.35$, $\mathrm{RA = 19^h31^m49^s.60}$, $\mathrm{Dec = -26^\circ34'34''}$, detected in the Hubble Cluster Lensing and Supernova Survey \citep[CLASH; ][]{Postman12}. The BCG of this cluster has a massive \Htwo~reservoir ($\sim 2 \times 10^{10} \rm M_{\odot}$), possibly feeding the strong star formation activity of the galaxy \citep[$\rm SFR \sim 250 \rm \, M_{\odot} \, yr^{-1} $;][]{Fogarty19}. This gas reservoir is detected as a tail-like structure extended in the northwest direction of the BCG for $\sim 30$ kpc in X-ray, optical, submillimeter, and infrared \citep{Ehlert11, Giacintucci14, Fogarty19, Ciocan21, Ghodsi24}. The BCG of MACS1931 contains an X-ray luminous AGN \citep{Ehlert11}.
The strong AGN feedback of the BCG has made two large X-ray cavities in the direction of east-west. Two X-ray bright ridges extend for almost 25 kpc to the north and south of the BCG. The north ridge is cooler (X-ray temperature of kT=$4.78 \pm 0.64$ keV), denser (peak of the density profile $\rm \sim 0.1 \, cm^{-3}$), and more metal-rich (Z=$\rm 0.53 \pm 0.11 \, Z_{\odot}$) compared to the rest of the cluster ($\rm \langle kT \rangle=5.87^{+1.30}_{-0.46}$ keV, $\rm \langle Z \rangle= 0.22 \pm 0.09 \, Z_{\odot}$), and is co-spatial with the CGM tail \citep{Ehlert11}. Very Large Array (VLA) observations of \citet{Giacintucci14}  reveal two radio sources at 1.4 GHz, one coincident with the BCG and another more compact source at a distance of $\sim 3''$ in the direction of the tail.
\citet{Ghodsi24} models the excitation of the CO and [CI] lines, tracing the relatively cold phase of gas, and shows that the tail gas is slightly colder and less dense ($\rm T_k=15.3_{-5.3}^{+15.6} $K, $\rm{\log{\frac{n_{H_2}}{[cm^{-3}]}}}=3.5_{-1.6}^{+2.2}$) than the interstellar medium (ISM) of the BCG ($\rm T_k=22.6_{-4.3}^{+8.3} $K, $\rm{\log{\frac{n_{H_2}}{[cm^{-3}]}}}=4.5_{-0.8}^{+1.3}$).
The nature of this multiphase gas reservoir that is located in a complex environment is not clear. 
In this work, we jointly analyze the distribution of cold and warm molecular hydrogen traced by Atacama Large Millimeter/Submillimeter Array (ALMA) and the James Webb Space Telescope (JWST), respectively, to investigate the spatial distribution, kinematics, and origin of the warm and cold molecular gas in the BCG and its CGM.

The details of the observations are explained in Section~\ref{sec:Observations}; the results of our analysis on MACS1931 are presented in Section~\ref{sec:Results}; Section~\ref{sec:Discussion} discusses the implications of this work; and Section~\ref{sec:Conclusion} summarizes our work.
Throughout this work, we take the systemic redshift of the BCG to be $z=0.352$ \citep{Postman12}. We assume a cosmology with parameters $\rm H_0 = 70~km~s^{-1}~Mpc^{-1}$, $\rm \Omega_m = 0.3$, and $\rm \Omega_{\Lambda} = 0.7$ in this work. We assume a factor of 1.36 in converting the hydrogen mass to the total mass including helium and heavy elements.

\begin{table*}[ht]
\centering
\caption{Properties of H$_2$ rotational lines detected in MACS1931 BCG.}
\label{tab:H2_lines}
\begin{tabular}{|ccccc|cccccc|}
\hline
Line & Transition & Rest $\lambda$ & $E_u/k$ & $A$ & MRS Ch. & $\text{FoV}_\text{tot}$ & $R$ & $\Delta v$ & PSF FWHM & RMS\\
 & & [$\mu$m] & [K] & [$10^{-11}\,\mathrm{s}^{-1}$] & & $[\text{arcsec}^2]$ & & [km\,s$^{-1}$] & [arcsec] & [$\mu$Jy\,pixel$^{-1}$] \\
\hline 
\Snine & J=11$\rightarrow$9 & 4.6947 & 10263 & 49000 & 1 & 122 & 3791 & 79 & 0.32 & $3.63$ \\
\Sfive & J=7$\rightarrow$5 & 6.910 & 4585 & 5880.0 & 2 & 146 & 3407 & 88 & 0.41 & $7.00$ \\
\Sone & J=3$\rightarrow$1 & 17.035 & 1015 & 47.6 & 4 & 259 & 1655 & 181 & 0.87 & $68.20$ \\
\hline
\end{tabular}
\tablecomments{Columns show molecular hydrogen transition properties (rest wavelength, upper-level energy, Einstein coefficient $A$) and observational parameters (MIRI MRS channel number, FoV for the total six pointing mosaic, spectral resolution $R$ and the corresponding value in velocity units $\Delta v$, PSF FWHM, and the typical root mean squared (RMS) flux per pixel) from JWST/MIRI MRS observations of MACS1931 BCG. The last four columns are all calculated at the observed wavelengths of the listed lines at the redshift of the BCG. }
\end{table*}

%%%%%%%%%%%%%%%%%%%%%%%%%%%%%%%%%%%%%%%%%%%%%%%%%%%%%%%%%%%%%%%%%%%%%%%%%%%%%%%%%%%%%%%%%%%%%%%%%%%%%%%%%%
\section{Observations}
\label{sec:Observations}

\subsection{JWST Observations}

We obtained Mid-Infrared Instrument (MIRI) Medium Resolution Spectrometer (MRS) observations of MACS1931 as part of JWST Cycle 2 GO program ID 3629 \citep[PI: A. Man, co-PIs: P. Andreani, L. Ghodsi; ][]{2023jwst.prop.3629M}. The observations were conducted on 2023-09-30 over a mosaic of six pointings (3 rows $\times$ 2 columns with field of view (FoV) per pointing from $3.2''\times 3.7''$ for channel 1 up to $6.6''\times7.7''$ for channel 4, designed to cover the full 30 kpc molecular gas tail detected by ALMA and the BCG core, with a four-point dither pattern to improve spatial sampling and mitigate bad pixels. The on-source integration time was 1154.417 seconds per pointing for a total of 6926.502 seconds on-source integration time. The spatial and spectral resolution values are shown in Table \ref{tab:H2_lines}. During the science observations, simultaneous MIRI imaging with the F2100W filter was conducted on an adjacent field to assist with astrometric alignment. A dedicated sky observation was performed for background subtraction (one pointing of 278 seconds), its location chosen to avoid bright sources at MIRI wavelengths (based on IRAC and WISE observations), and to enable simultaneous MIRI imaging of the BCG with the same F2100W filter.

The observations were carried out using the full set of four integral field units (IFUs, channels 1 to 4), utilizing the MRS medium grating and thus cover the spectral ranges of [5.66--6.63 $\mu$m, 8.67--10.13 $\mu$m, 13.34--15.57 $\mu$m, 20.69--24.48$\mu$m], targeting the \Sone~line in channel 4, \Sfive~line in channel 2, and \Snine~in channel 1. $\rm H_{2}\,S(j)$ is a short notation of $\rm 0-0\,H_2\, S(v=0\rightarrow0, J=j+2\rightarrow j)$ for the pure rotational transitions of molecular hydrogen in the S-branch, showing no change in the vibrational quantum number $\Delta v =0$ and an increase in the rotational quantum number $\Delta J = -2$. Table \ref{tab:H2_lines} summarizes the important properties of these \Htwo~emission lines\footnote{Adopted from Gemini observatory resources (\url{https://www.gemini.edu/observing/resources/near-ir-resources/spectroscopy/important-h2-lines})}.
Both the spectral and spatial resolution vary with wavelength. The spectral resolution, characterized by the resolving power \( R = \lambda/\Delta \lambda = c/\Delta v \), can be calculated using Equation 1 of \citet{Jones2023}, while the spatial resolution, represented by the full width half maximum (FWHM) of the PSF, follows Equation 1 of \citet{2023AJ....166...45L}, assuming a linear fit. The details of the calibration are presented in \autoref{sec:jwst_data_cal}.

\subsection{ALMA Observations}

We use archival ALMA Band 6 observations of MACS1931 (project ID: 2017.1.01205.S, PI: M.Postman), presented in \citet{Fogarty19} and re-analyzed in \citet{Ghodsi24}. We refer the reader to \citet{Fogarty19} for a detailed description of the observations and the calibration procedure. This dataset detects \COthree~in the BCG of MACS1931 and its CGM tail. For comparing with the JWST \Htwo~observations, we use ALMA Band 6 observations at the frequency range of [238.524 -- 258.509] GHz covering \COthree, as it has the highest signal-to-noise ratio (SNR) among the ALMA observed spectral lines.

The ALMA Band 6 calibrated measurement sets were retrieved from the science ready data product initiative \citep[SRDP\footnote{\href{https://science.nrao.edu/srdp}{https://science.nrao.edu/srdp}};][]{Lacy20}, calibrated using the standard ALMA pipeline in \texttt{CASA 6.5.4-9} \citep[Common Astronomy Software Applications package;][]{McMullin07}.
We clean and image all 12 m array and ACA measurement sets simultaneously using task \texttt{tclean} of CASA in interactive mode with a threshold of 0.05 mJy, using a pixel size of $0.075''$, a velocity channel width of $25 \, \rm km\,s^{-1}$, and a natural weighting, following \citet{Fogarty19}. To combine the measurement sets having slightly different frequency tunings, we exclude 21 channels from the edges of each spectral window.
 In the image plane, we use the CASA task \texttt{imcontsub} to fit the continuum with a zeroth-order polynomial to the line-free channels at [254.67--255.16] GHz and [256.67--258.29] GHz and subtract it from the image, which gives a better residual compared to the continuum subtraction in the uv plane. The final beam has a major axis of $0.91''$ and a minor axis of $0.76''$, on par with or larger than the MIRI MRS PSF (see \autoref{tab:H2_lines}). The root mean square (RMS) of the line-free channels in the data cube is 0.28 mJy\,beam$^{-1}$.

%%%%%%%%%%%%%%%%%%%%%%%%%%%%%%%%%%%%%%%%%%%%%%%%%%%%%%%%%%%%%%%%%%%%%%%%%%%%%%%%%%%%%%%%%%%%%%%%%%%%%%%%%%
\section{Results} 
\label{sec:Results}

\subsection{Emission Lines}
\label{ssec:emission_lines}

\begin{figure*}
%\hspace{2cm}
%\begi{center}
\includegraphics[width=18cm]{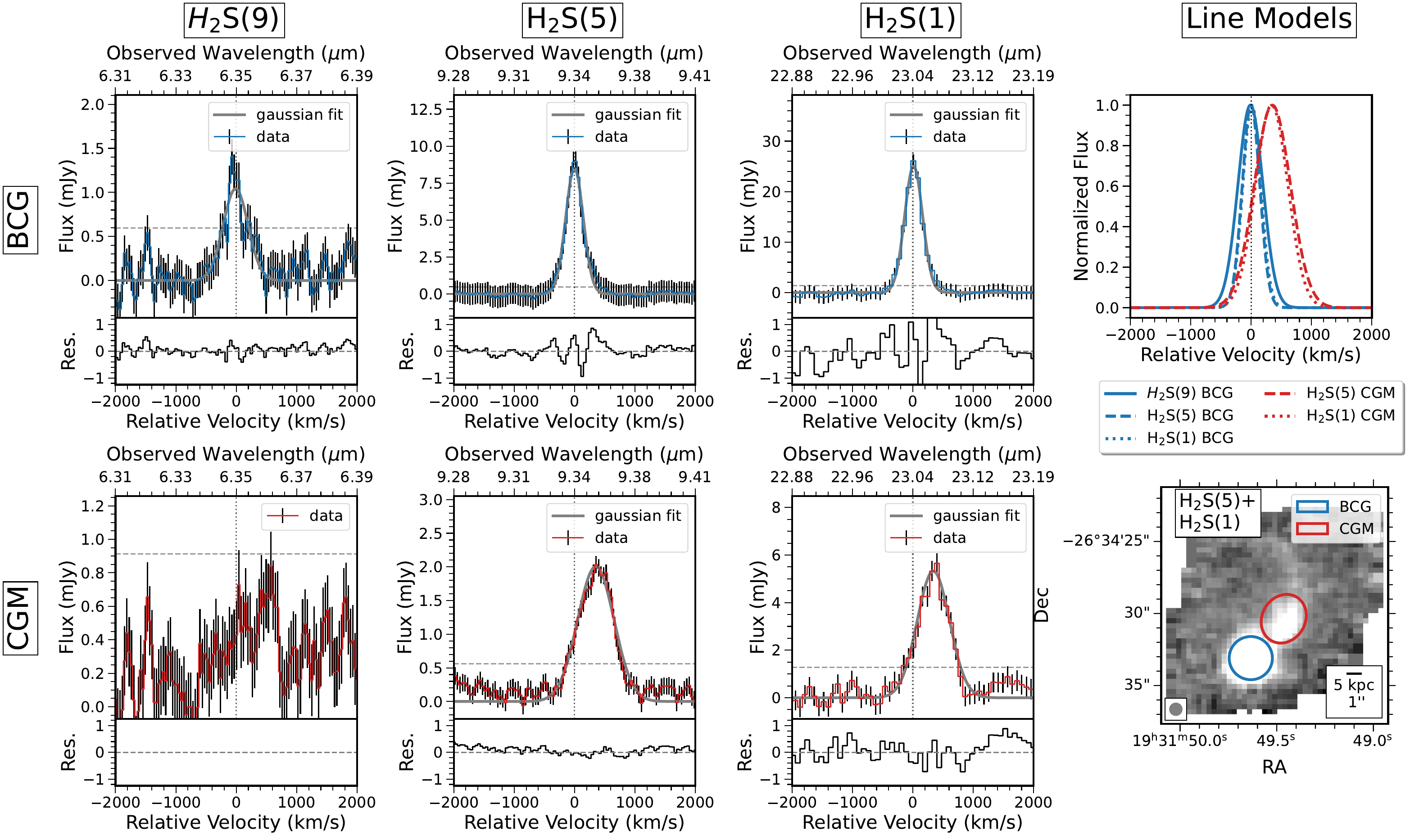}
\caption{Continuum-subtracted spectra toward the three \Htwo~S(9), S(5) and S(1) lines in the first three columns. Spectra are extracted from two apertures centered on the BCG (top row) and the CGM (bottom row), respectively. The bottom-right panel illustrates the spatial extraction regions of the BCG and CGM, overlaid on a stacked moment zero map of \Sone~and \Sfive. The grey circle in the bottom left of the panel illustrates the size of the MIRI PSF FWHM of the lower spatial resolution map, \Sone, to which the \Sfive~map was convolved to match. The extracted spectra are shown as colored step plots as a function of velocity relative to the systemic redshift. The Gaussian models (solid grey) are shown for all lines other than \Snine~in the CGM as there is no distinguishable line profile to fit. The 1$\sigma$ uncertainties on the flux are shown as black vertical bars (see \autoref{sec:ferr_est} for more information). The dashed horizontal and dotted vertical grey lines represent the $3\sigma$ flux upper limits and systemic velocity of the BCG, respectively.  
The 3$\sigma$ flux upper limits are computed as three times the RMS noise (per velocity channel) over the velocity range [$-2500$,$+2500$]$~\mathrm{km\,s^{-1}}$, excluding the interval used for line flux integration (defined as the mean velocity of the Gaussian fit $\pm 3\times\mathrm{FWHM}$ of the fitted profile). Residuals of the fits in mJy are displayed beneath each spectral plot. The top-right panel presents a normalized comparison of all line models, with BCG extractions in blue and CGM extractions in red in order to illustrate the velocity offset of \Htwo~emission line from the CGM tail. 
}
 \label{fig:H2line_spec1}
% \end{center}
\end{figure*}

\begin{table*}[ht]
\centering 
\caption{Molecular hydrogen rotational lines observed for MACS1931.}
\label{tab:MACS_lines}
\hskip-2.0cm\begin{tabular}{ |lccccccc| } 
\hline
Region & Transition& Velocity Offset & Amplitude  & FWHM  & Integrated Flux & Intrinsic Integrated Flux &$\rm SNR_{peak}$  \\
 & &[$\text{km}\,\text{s}^{-1}$] & [mJy] & [$\text{km}\,\text{s}^{-1}$] & [$10^{-18} \text{W}\, \text{m}^{-2}$] & [$10^{-18} \text{W}\, \text{m}^{-2}$]  &\\
\hline
BCG &\Sone & 13.0 $\pm$ 3.4 &25.0 $\pm$ 0.5 &301 $\pm$ 9 & 4.11 $\pm$ 0.08 & 5.00 $\pm$ 0.12 & 54.6\\
    &\Sfive & -1.8 $\pm$ 4.1 &8.5 $\pm$ 0.2 & 332 $\pm$ 12 & 3.41 $\pm$ 0.08 & 3.82 $\pm$ 0.09 & 55.5 \\
    &\Snine & -11 $\pm$ 11 & 1.0 $\pm$ 0.1 & 481 $\pm$ 33  & 0.86 $\pm$ 0.04 & 1.03 $\pm$ 0.05 & 5.3\\
\hline
CGM & \Sone & 340 $\pm$ 7 & 5.3 $\pm$ 0.1 & 605 $\pm$ 18 & 1.48 $\pm$ 0.04& 1.81 $\pm$ 0.09 & 12.5\\
    &\Sfive & 352 $\pm$ 4 & 2.0 $\pm$ 0.1 & 702 $\pm$ 11 & 1.58 $\pm$ 0.02 & 1.78 $\pm$ 0.02 & 10.8\\
    &\Snine & - & - & -  & $3\sigma$ upper limit = 0.43 & - & -\\

\hline
\end{tabular}
\tablecomments{The emission line velocity offset, amplitude, line FWHM, integrated observed flux, intrinsic flux (corrected for extinction) with uncertainties, and the $\rm SNR_{peak}$ (amplitude divided by RMS per velocity channel integrated over apertures) of the fits are presented. The FWHM values have been corrected for instrumental broadening using spectral resolutions calculated using Equation 1 of \citet{Jones2023}. The integrated fluxes are corrected for dust extinction (see section \ref{ssec:emission_lines}).}
\end{table*}

We extract the spectra of the detected rotational molecular hydrogen lines in the JWST data from a circular aperture centred on the BCG (center: [19h$\,$31m$\,$49.63s, -26$^\circ\,$34$'\,$33.10$''$], with radius $1.5''= 7.5$ kpc), and an elliptical aperture on the tail defined based on the \Htwo~and CO intensity contours (center: [19h$\,$31m$\,$49.46s, -26$^\circ\,$34$'\,$30.36$''$], with major axis of $3.5''$, minor axis of $3''$, position angle of $60^\circ$). These apertures are chosen to maximize SNR of the \Htwo~and CO(3-2) emission. 

To model and subtract the underlying continuum in our data, we apply Cubeviz's pixel-by-pixel fitting procedure \citep{jdadf_developers_2024_12797328} to fit a linear 1D model to each pixel in the cube, generating a 3D continuum cube. This approach is sufficient for creating moment maps of the lines of interest; however, for calculating spectrally integrated line fluxes in 1D, we found that a 5th-order polynomial provided a better continuum fit for these lines. As demonstrated by \citet{Fogarty19}, dust continuum emission is detected in the CGM region using ALMA Band 6 and 7 data. Due to the saturation of the BCG in the MIRI photometry at 21 $\mu$m (F2100W) we cannot draw useful information about the MIR continuum within the tail with this dataset.

The continuum-subtracted spectra are shown in Figure \ref{fig:H2line_spec1}. We fit the emission lines with a single Gaussian centered on the systemic redshift of MACS1931 and compare the normalized model profiles in the upper-right-most panel. The best-fitting parameters are summarized in Table \ref{tab:MACS_lines}. 
The integrated observed fluxes are obtained by summing fluxes over the line FWHM. We propagate uncertainties on the fits using a Monte Carlo approach, where we use 10000 realizations of the spectrum perturbed by the errors (assumed to follow a Gaussian distribution) and fit each spectrum. The reported fit parameters and errors refer to the average and standard deviation of the 10000 realizations, respectively.
The \Snine~line is only detected in the BCG, but not in the tail.  We estimate the $3 \sigma$ upper limit for the \Snine~line in the CGM tail by computing $\sigma = \mathrm{RMS \sqrt{N_{ chan}}  \delta v}$, where $\mathrm{RMS}=0.31\rm \, \mathrm{mJy}$ is the noise per channel of the spectrum, $\mathrm{N_{chan}} = 2\times \mathrm{FWHM/\delta v}$ is the number of line channels, FWHM is the line full width at half maximum, and $\mathrm{\delta v=37 \,  km \, s^{-1} }$ is the channel width, which corresponds to  7.48 GHz. We assume the FWHM of the \Snine~line to be the same as that of \Sfive~line in CGM and calculate the upper limit to be $4.3 \times 10^{-19}\, \rm Wm^{-2}$.

In addition, we calculate extinction correction factors to the observed line fluxes using the attenuation curve derived from the CAFE spectral fitting code\footnote{\url{https://github.com/GOALS-survey/CAFE}} (\citealt{Diaz-Santos2025}; see \citealt{Lai2022,Jones24} for use case scenarios). We apply these corrections to obtain intrinsic line fluxes in \autoref{tab:MACS_lines}. CAFE corrects for dust extinction using the OHMc (Overlapping H\textsc{ii} region and Molecular cloud) extinction curve \citep{Ossenkopf1992, 2008ApJ...678..729S}. For the CAFE fits, we choose a baseline dust model that includes both emission and absorption from warm dust ($100$–$400$\,K) and assumes a foreground screen geometry. In CAFE, emission line attenuation is governed by the optical depth of the warm dust component. The best-fit parameters of the model are an optical depth of $\tau_\text{warm} = 1.20\pm0.06$ and a dust temperature of $T_\text{warm} = 171\pm17\,\text{K}$.  We have accounted for the uncertainties from the CAFE extinction curve in the intrinsic integrated flux values by performing a similar Monte Carlo error propagation mentioned above with 10000 iterations based on the CAFE spectrum fits.

The linewidth of emission lines is related to the turbulence and the distribution of velocities of gas clumps in the line of sight. The FWHM of the \Htwo~rotational lines range from 301 $\rm km \, s^{-1}$ to 481 $\rm km \, s^{-1}$ for the BCG and from 605 $\rm km \, s^{-1}$ to 702 $\rm km \, s^{-1}$ for the CGM tail after correcting for the MIRI spectral resolution (Table \ref{tab:MACS_lines}). This might be due to the more disturbed medium of the CGM or the presence of multiple gas clumps in the line of sight or the effect of inclination. 
We have measured the line width of \COthree~over the same BCG and CGM apertures used for the \Htwo~lines and corrected the values for the rest-frame spectral resolution of the ALMA Band 6 observations ($18.1 \, \rm km \, s^{-1}$). Consequently, the intrinsic line width of the warm \Htwo~is on average $\sim 120 \, \rm km \, s^{-1}$  higher than the \COthree~line FWHM. Part of this is due to the temperature difference in the gas probed, and a part might be due to the different geometry of the warm and cold clouds \citep{Ubler19}.

\begin{figure*}

\includegraphics[width=18cm]{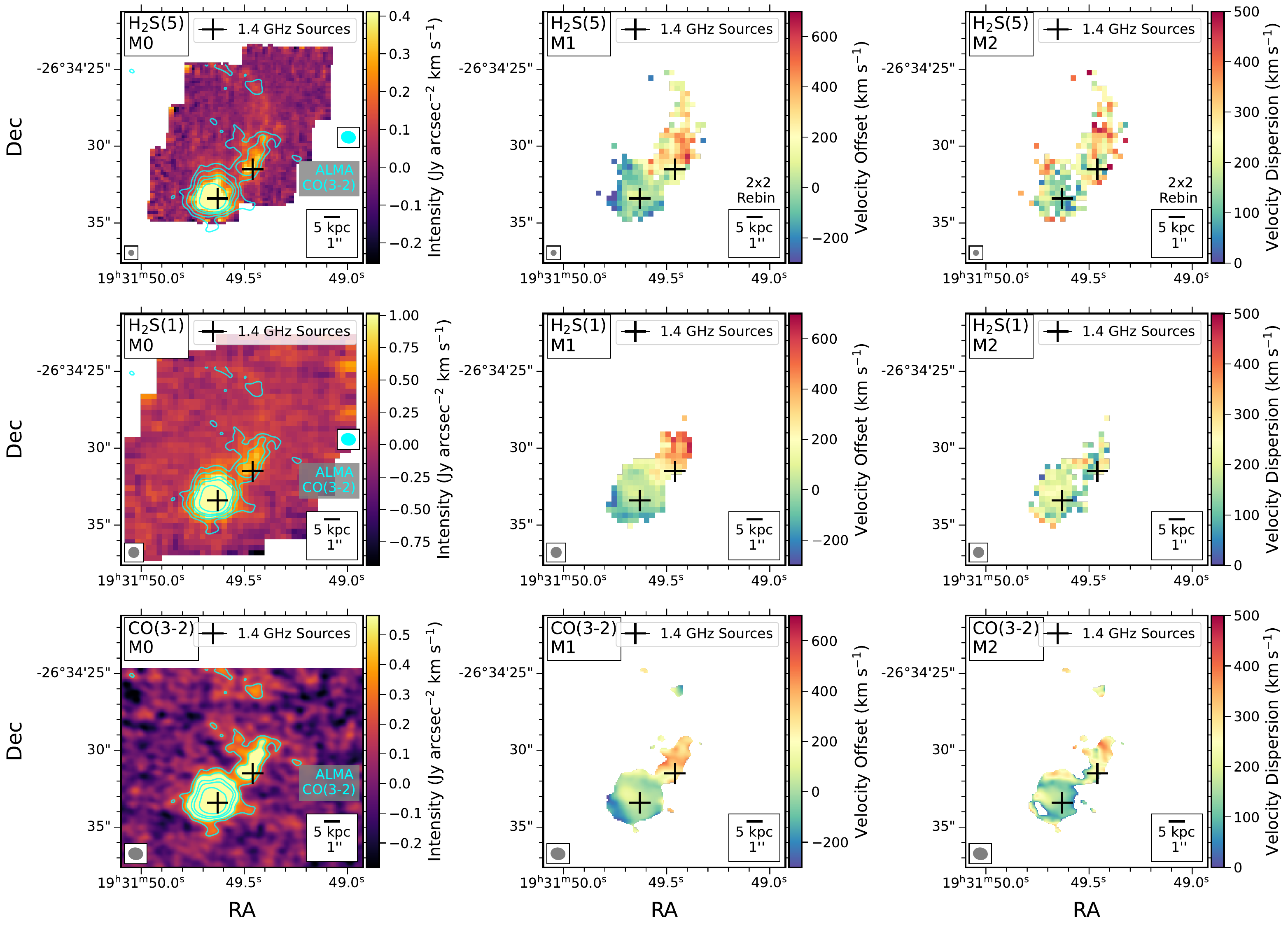}
\caption{Moment maps of the molecular gas tracers in MACS1931 BCG, showing similar spatial distribution and kinematics for warm and cold gas. \textit{Left}: Intensity maps of the \Sfive, \Sone, and \COthree~emission lines from top to bottom, respectively. ALMA \COthree~contours are overplotted in cyan shown at $[0.5,1,2,3.6]\sigma$ with $\sigma=0.46 \, \rm Jy \, arcsec^{-2} \, km \, s^{-1}$. \textit{Middle}: Velocity offset maps of the same lines. \textit{Right}: Velocity dispersion maps of the same lines. The moment one and two maps only show pixels with SNR$>3$ in the moment zero map. The velocities are defined relative to the systemic redshift of the BCG. The grey circle in the bottom left of the \Htwo~plots illustrates the size of the MIRI PSF FWHM, and the cyan ellipse is the synthesized beam of ALMA \COthree~line. The grey ellipse in the CO maps in the last row shows the ALMA synthesized beam. The black crosses indicate the centroid of the 1.4 GHz emission \citep{Giacintucci14}. The black scale bar indicates 5 proper kpc at the systemic redshift of the BCG or equivalently 1 arcsecond on the sky.  We spatially bin the moment one and two maps of \Sfive~by $2\times2$ pixels in order to improve SNR.}
 \label{fig:mom1-2}
\end{figure*}

Figure \ref{fig:mom1-2} shows similar spatial distribution and kinematics for the warm and cold molecular gas \Sfive, \Sone, and \COthree. Both \Sfive~and \Sone~show that the tail is redshifted by approximately $250-350 \, \rm km \, s^{-1}$, similar to that seen in \COthree. A blue-shifted arc in the southeast of the BCG is visible with a velocity of approximately $-150 \, \rm km \, s^{-1}$ to $-250 \, \rm  km \, s^{-1}$. 
This figure shows that the \Htwo-traced warm gas shares similar kinematics with the CO-traced cold gas,  with a redshifted CGM tail. The slightly different properties of the \Sone~moment two map may be intrinsic or arise from low SNR.

\subsection{Single-Temperature Modeling}
\label{sec:Tex}

If we assume that the warm \Htwo~gas is in local thermodynamic equilibrium (LTE) and is optically thin, the rotational level population follows the Boltzmann distribution, as done in \citet{Rigopoulou02, Ogle10, Guillard12, Pereira22, Dasyra24}. In this case, the ratio of the column density of a transition, $N_i$, to the total column density, $N_{\mathrm{tot}}$, would be

\begin{equation}
    \frac{N_i}{N_{\mathrm{tot}}} = \frac{g_i}{Z_{T_{ex}}} \times \exp \left( -\frac{T_i}{T_{ex}} \right)
    \label{eq:Tex}
\end{equation}
\\
where $g_i$ is the statistical weight of the $\rm i^{th}$ transition; $T_i$ is the energy level of the $\rm i^{th}$ transition in units of Kelvin; and $T_{ex}$ is the excitation temperature of the lines which is the gas temperature at LTE; and $Z_{T_{ex}}$, the partition function at the excitation temperature, equals 
\begin{equation}
    \sum_{\mathrm{even} \, j}^{\infty}(2j+1)\, \exp(-E_j/kT_{\rm ex}) \, + \, \sum_{\mathrm{odd} \, j}^{\infty}3(2j+1)\, \exp(-E_j/kT_{\rm ex})
\end{equation}

Here we have assumed that ortho-to-para ratio is 3 for gas with temperature $>250$ K \citep{Burton92, Rigopoulou02, Togi16}. The column density of each transition is related to the transition flux, $F_i$ in units of $\rm Wm^{-2}$ through

\begin{equation}
    N_i = \frac{F_i}{A_i h \nu_i} \times \frac{4 \pi}{\Omega}
\label{eq:Ni}
\end{equation}

where $A_i$ is the Einstein coefficient; $h$ is the Planck's constant; $\nu_i$ is the transition frequency; and $\Omega$ is the solid angle of the aperture in units of steradian. Finally, we calculate the total mass of warm molecular hydrogen using $M_{H_2}=N_{\mathrm{tot}}\Omega d^2 m_{H_2}$ with $d$ and $m_{H_2}$ being the angular diameter distance to the object and the mass of hydrogen molecule, respectively.

\begin{figure*}
\hspace{-1cm}
%\begi{center}
\includegraphics[width=20cm]{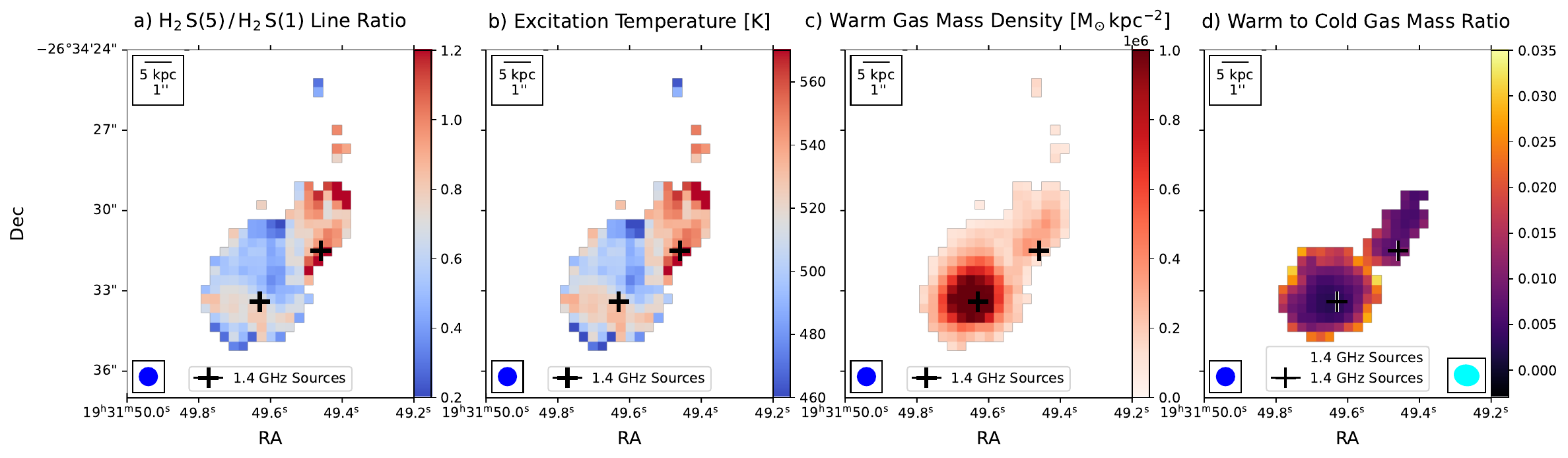}
\caption{\textit{(a)} The line ratio map of \Sfive~to \Sone, calculated using the unbinned extinction-corrected intensity maps with extinction corrected. \textit{(b)} The excitation temperature calculated under the LTE assumption using the \Sfive~and \Sone~lines through Equation \ref{eq:Tex}. \textit{(c)} Surface density of warm molecular gas mass. \textit{(d)} Ratio of the warm-to-cold molecular gas mass. This map is calculated using the warm mass map shown in the left panel of Figure \ref{fig:Tex} and cold gas masses calculated based on ALMA CO(3-2) emission. As in Figure \ref{fig:mom1-2}, the black crosses are the VLA 1.4 GHz radio sources. The black scale bar indicates 5 kpc in space or equivalently 1 arcsecond on the sky. The dark blue circle shows the size of the MIRI PSF FWHM at the wavelength \Sone~and the cyan ellipse shows the ALMA Band 6 beam.}
 \label{fig:Tex}
% \end{center}
\end{figure*}

Figure \ref{fig:Tex} shows the outcome of applying the above-described method on the two hydrogen lines best detected in our data. Panel (a) shows the line flux ratio of \Sfive~to \Sone~calculated using the unbinned moment zero maps of the two lines. For this step, we have convolved the \Sfive~map with a Gaussian kernel to match the PSF of the \Sone~map, essentially smoothing by a kernel with FWHM=$\rm \sqrt{FWHM_{ch4}^2 - FWHM_{ch2}^2}=0.77''$, and then resampled it to the pixel scale of the \Sone~map. We have applied extinction correction factors to the moment maps before calculating the line ratio map to account for the difference in dust extinction at different MIRI channels. 
Panel (b) of Figure \ref{fig:Tex} shows the calculated excitation temperature using these two lines. The excitation temperature in the BCG is $515.6 \pm 0.8$ K on average. This temperature is a weighted average based on the excitation temperature uncertainties derived from equations \ref{eq:Tex} and \ref{eq:Ni}, and the uncertainty is the weighted average of the uncertainties over the region. The temperature distribution within the BCG is a bimodal distribution with a scatter of 18 K. A part of the BCG closer to the tail is slightly colder ($503 \pm 1$ K) and the south-east blue-shifted arc (see section \ref{ssec:emission_lines}) is slightly warmer than the center ($525 \pm 1$ K). The tail is on average warmer than the BCG ($535.2\pm 1.9$ K ) with a scatter of 22 K. The excitation temperatures of the BCG and CGM tail are higher ($\sim 522$ K and $\sim 546$ K, respectively) when the extinction correction from CAFE is not applied. This is because the extinction factor is lower in channel 2 than in channel 4, resulting in a higher \Sfive/\Sone~ratio in the uncorrected case. Note that we use the same extinction correction factor across the entire system.

The surface density of warm molecular gas mass is presented in Figure \ref{fig:Tex}c. The BCG and the tail contain $(1.3 \pm 0.4) \times 10^8$ \Msun~and $(3.3 \pm 0.3) \times 10^7$ \Msun~of warm gas respectively with an average gas surface density of $(6.3 \pm 0.1) \times 10^5 \, \rm M_{\odot}\,kpc^{-2}$ and $(2.0 \pm 0.1) \times 10^5 \, \rm M_{\odot}\,kpc^{-2} $ for the BCG and the CGM tail. The measured warm gas mass is then used with the cold gas mass map of ALMA \COthree~to calculate the warm-to-cold gas mass ratio, presented in Figure \ref{fig:Tex}d. We use the \COthree~line here since this line has the best signal-to-noise ratio among all CO lines and the tail structure is clearly visible in the moment maps of this line, which makes it easier to separate between BCG and the tail. In order to make this map, we have converted the \COthree~fluxes to \COone~ using the CO spectral line energy distribution (CO SLED) of MACS1931 BCG and CGM determined in \citet{Ghodsi24},  using a conversion factor for the velocity-integrated fluxes of $8.7$ for the BCG and 6.0 for the CGM tail. Then we calculate the cold molecular gas mass in each pixel of the 2D image assuming a conversion factor of $\alpha_{CO}=M_{mol}/L^{\prime}_{CO(1-0)} = 0.8 \, \rm M_{\odot}/[K\,km\,s^{-1}\,pc^2]$ \citep{Bolatto13}. Using \COthree~instead of \COone~gives us a total molecular gas mass of $(1.2 \pm 0.1) \times 10^{10}$ \Msun~for the whole system which is in good agreement with the \COone-derived mass of molecular gas, $(1.9 \pm 0.3) \times 10^{10}$ \Msun, presented in \citet{Fogarty19}. 

The rightmost panel of Figure \ref{fig:Tex} shows that the ratio of the warm-to-cold gas mass of the CGM tail ($1.9\%\pm 0.3\%$) is equal to the BCG ($1.4\%\pm 0.2\%$) within error bars. The total mass of warm molecular gas assuming LTE is $(1.6 \pm 0.3) \times 10^8$ \Msun~in the whole system, which is $\sim0.9\%$ of the cold gas mass estimated by \citet{Fogarty19}, assuming $\alpha_{CO}= 0.8 \, \rm M_{\odot}/[K\,km\,s^{-1}\,pc^2]$. 

\subsection{Continuous Temperature Modeling}
\label{Togi}

 In the regions where non-thermal heating mechanisms play a role, more sophisticated temperature modeling than the single-temperature fits may be required to capture their effect on the temperature distribution of the gas. In this section, we instead use a continuous temperature model to explain the observed \Htwo~emission.

\subsubsection{Model Description and Formalism}
 
\citet{Togi16} models the hydrogen rotational excitation with a temperature power law model to a sample of galaxies under thermal equilibrium, which allows a broader range of temperatures for the gas, rather than a single temperature or two-component temperature models. The fundamental assumption of the model is
$ dN = mT^{-n} dT $;
where $dN$ is the column density of the $\rm H_2$ gas with temperature between $T$ and $T+dT$, $m$ is a normalization factor, and $n$ is the power law index. Based on this assumption, the column density of gas at upper energy level j is derived as

\begin{equation}
N_j = \int_{T_l}^{T_u}  \frac{g_j}{Z(T)} \exp(\frac{-E_{j \rightarrow j-2}}{k_BT}) m T^{-n} dT    
\end{equation}

where $g_j$ is the degeneracy factor of the upper level $j$, $Z(T)$ is the partition function, $E_j$ is the energy of level $j$, $k_B$ is the Boltzmann constant, and [$\rm T_l, T_u$] show the temperature limits where this assumption is valid (see \citealt{Togi16} for more details).
Using this formalism, they model the excitation of \Htwo~rotational lines and infer their model parameters ($\rm n, T_l, T_u$). We need the detection of more than two emission lines to constrain this distribution properly.

We apply the \citet{Togi16} model to the \Htwo~measured fluxes for the MACS1931 BCG where we detect all three \Sone, \Sfive, and \Snine~lines. We fix the upper temperature of the model to $T_u=5000$ K since the variations of this parameter above this value does not impact the fit, and then we fit the other two parameters $ n,\, T_l$. In regions with high warm gas mass fraction, the excitation diagram is flatter (lower n) and $T_l$ could be interpreted as the lowest possible temperature needed to explain the \Htwo~excitation. We add a Markov chain Monte Carlo (MCMC) ensemble sampler using the Python package \texttt{emcee}\footnote{\href{https://github.com/dfm/emcee}{https://github.com/dfm/emcee}}\citep{Foreman13} to find the best model and uncertainties, assuming a flat prior with $2<n<8$ and $10<T_l<1000$ K. These ranges are chosen based on typical values found in this analysis for other galaxies. 
We maximize the model likelihood $P={\rm exp}(-\chi^2/2)$, with  $\chi^{2}$=$\Sigma_i
(f_{{\rm  obs}(i)}  -  f_{{\rm  model}(i)})^2/\sigma_i^{2}$ where $f = \ln{(\frac{N_u/g_u}{N_3/g_3})}$ is the excitation function for the model and observations and $\sigma$ is the observational uncertainty. This function characterizes the relative population of \Htwo~rotational levels and provides insight into the excitation conditions of the molecular gas \citep{Rigopoulou02, Roussel07}. The MCMC sampling approach gives us the highest likelihood model parameters for the BCG. 

\begin{figure}
\hspace{0cm}
%\begi{center}
\includegraphics[width=9cm]{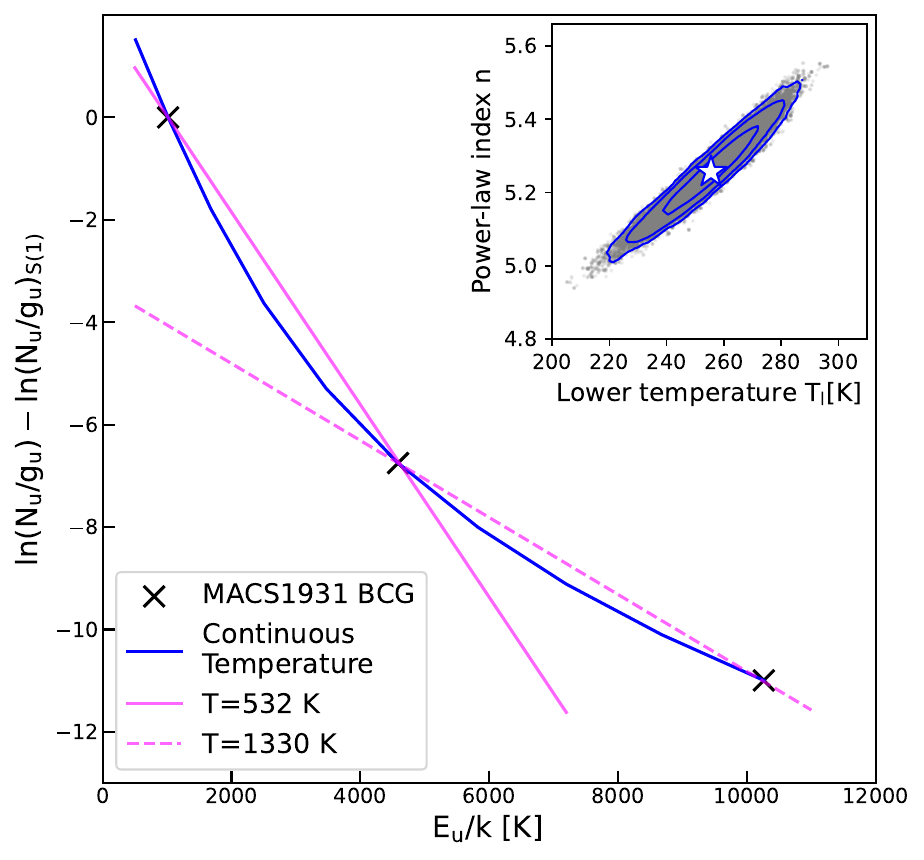}
\caption{The \Htwo~excitation diagram for the BCG region of MACS1931. The crosses show the observed values. The error bars are very small compared to the marker size. The blue curve shows the continuous temperature fit using the \citet{Togi16} model. The magenta lines show single temperature fits to the \Sone-\Sfive~transitions (solid) and the \Sfive-\Snine~transitions (dashed) and the best-fitting temperatures are shown in the legend. The inset shows the distribution of the continuous temperature model parameters drawn from the MCMC sampling. The best model is shown with a star, and the contours show the [68, 95, 99] percentiles.}
 \label{fig:TogiSmith}
% \end{center}
\end{figure}

Figure \ref{fig:TogiSmith} shows the excitation function and the MCMC results for the BCG, calculated using the parameters of the models with highest likelihood in the MCMC sampling, $\rm T_l=256 \pm 11$ K and $\rm n=5.26 \pm0.08$ and compares that to the single-temperature fit. The solid magenta line calculated using the integrated line fluxes is similar to the spatially-averaged temperature presented in Section \ref{sec:Tex} calculated using the single-temperature fitting. The difference between these two values originates from the spatial averaging process. The difference between the BCG masses measured using the single temperature and continuous temperature methods is approximately 6\%. While this difference is not significant, we can better constrain the temperature distribution when more MIR \Htwo~lines are detected, as we expect for the approved Cycle 4 observations \citep{2025jwst.prop.8582M}.

\citet{Togi16} has done this analysis for the Spitzer Infrared Nearby Galaxies Survey (SINGS), which contains various types of galaxies, including luminous infrared galaxies (LIRGS). Their power-law index $n$ for the SINGS sample ranges from 3.79 to 6.39 with an average of $4.84 \pm 0.61$. The power-law index of the MACS1931 BCG is thus comparable to the SINGS galaxies. It is noteworthy that with this data for MACS1931 BCG, we can trace the warm gas with temperature higher than $\rm T_l=256$ K, which is higher than the average model $\rm T_l=81$ K for the SINGS galaxies with $\rm H_2 \, S(0)$ data. Consequently, the calculated warm-to-cold gas mass ratios only represent a part of the total warm gas and are lower than the SINGS galaxies ($2\% - 11\%$) of \citet{Togi16}.

\subsubsection{Total gas mass estimate}

Following \citet{Togi16}, we calculate the total warm molecular gas mass of the BCG to be $2.3 \times 10^8\,$\Msun~in the temperature range of $\rm (T_l - T_u) = (256-5000)$ K. The continuous temperature model can be extrapolated to lower temperatures to include the cold gas mass when estimating the total molecular hydrogen mass. Assuming the average extrapolation temperature for normal star-forming galaxies of $\rm T_l^*=49$ \citep{Togi16}, the total molecular gas mass of the MACS1931 BCG would be $(7.8\pm 0.1) \times 10^{10}$\Msun. Using ALMA observations of CO emission lines, \citet{Fogarty19} derives a total molecular mass of $(9.4 \pm 1.3) \times 10^{10}\,$\Msun~assuming a galactic CO-to-\Htwo~conversion factor. This comparison shows that the total gas mass estimated from the continuous temperature model extrapolation is in good agreement with the total mass predicted from CO.
However, as shown in \citet{Ghodsi24}, MACS1931 BCG has a CO-SLED similar to U/LIRGs and due to the lower CO-to-\Htwo~conversion factor $\rm \alpha_{CO}=0.8$\,\Msun$\rm  (K\,km\,s^{-1}\,pc^2)^{-1}$, a higher average extrapolation temperature of $\rm T_l^*=80$ is derived for these galaxies in \citet{Togi16}. Assuming this extrapolation temperature instead, the total molecular gas mass of MACS1931 BCG is $(9.9\pm 0.2) \times 10^{9}$\Msun. Instead, \citet{Fogarty19} calculates a total molecular gas mass of $(1.9 \pm 0.3) \times 10^{10}\,$\Msun~assuming the LIRGs conversion factor, meaning that $\sim 0.92 \times 10^{10}\,$\Msun~or 48\% of the total mass is not explained by the extrapolation.

%%%%%%%%%%%%%%%%%%%%%%%%%%%%%%%%%%%%%%%%%%%%%%%%%%%%%%%%%%%%%%%%%%%%%%%%%%%%%%%%%%%%%%%%%%%%%%%%%%%%%%%%%%
\section{Discussion}
\label{sec:Discussion}

In this work, we investigate the warm and cold molecular gas content of the BCG of the MACS1931 cluster and a part of its CGM using JWST \Htwo~rotational lines and ALMA \COthree~line. We measure the \Sone, \Sfive, and \Snine~fluxes and model the excitation temperatures. Our findings show that the cold and warm phases of \Htwo~ are co-spatial in the BCG and the CGM tail, and the CGM tail is slightly warmer than the BCG in the warm phase. In this section, we discuss the different physical processes and heating mechanisms that may be responsible for explaining the line ratios observed in this system.

\subsection{Warm-to-cold gas mass ratio}

We here contextualize our findings of the warm and cold molecular gas towards the MACS1931 BCG system with findings of nearby galaxies, as reported in JWST and Spitzer studies. The MACS1931 BCG is a complex object since it is the BCG of a massive cool-core cluster, it has an AGN and an active starburst, and our observations not only capture the circumnuclear emission, but also spatially resolve the extended reservoir beyond the BCG. We compare our results with various cases from the literature including AGN-host galaxies, starburst galaxies, and interacting systems. For this purpose, we compare two important parameters measured in these works: (i) Warm-to-cold gas mass ratio, which is related to the energy input from various physical processes. High warm-to-cold gas mass ratios are seen in regions undergoing energetic events where more gas is excited to above 100 K, resulting in emission in the IR. Normal star-forming galaxies have a typical warm-to-cold gas mass ratio of $\sim 1 \%$ \citep{Togi16}. (ii) Power-law index of the continuous-temperature modeling, which is related to the temperature distribution in the warm phase. A low power-law index shows a flat slope in the \Htwo~excitation diagram, corresponding to a large fraction of gas having high temperatures. Low power-law indices are seen in shock-heated regions, while high indices are common in star-formation dominated regions. For instance, in the SINGS sample of \citet{Togi16}, the average power-law index value for star-forming galaxies is $5.16 \pm 0.36$, while this value is $4.46 \pm 0.44$ for the AGN host galaxies \citep{Riffel25}.

AGN host galaxies may exhibit strong \Htwo~emission as a result of energetic processes such as turbulence and shocks occurring in their environments.
\citet{Reefe25} studies the centermost 50 kpc of the Phoenix cool-core cluster around the BCG, which is an AGN host at z=0.6. Phoenix BCG is the only cool-core BCG with MIR data other than MACS1931. They report a power-law index of $4.8 \pm 0.1$ for the gas in their whole FoV, in agreement with the value estimated for shock-heated regions. However, \citet{Reefe25} argues that the \Htwo~emission could be explained by the strong star formation of the BCG and there is no need for extra heating mechanisms. Three other nearby AGN host galaxies are studied in \citet{Costa-Souza24} and \citet{Riffel25} with JWST, where they detect multiple \Htwo~emission lines in the inner region of these galaxies. These works report a flatter power-law index for the \Htwo~excitation in kinematically disturbed regions compared to the virially dominated regions, indicating a major role of shocks in the kinematically disturbed regions.
\citet{Ogle24} detects warm \Htwo~gas with Spitzer Infrared Spectrograph (IRS) in the inner region of the AGN-host M58 galaxy. This work reports a power-law index of $n=4.32-5.44$ with a warm-to-cold gas mass ratio of $\sim 2\%$, similar to MACS1931 BCG. They suggest that the warm \Htwo~excitation results from shocks and turbulence from the radio jet.
Recent JWST observations of the central region of six quasars reveal a warm-to-cold gas mass ratio close to that of MACS1931, ranging from $0.4\%$ to $2\%$, originating from star formation, jet-induced shocks, turbulence, and UV heating \citep{Herrero24, Ramos25}.

Starburst galaxies might show different properties if their star formation is their dominant excitation mechanism. \citet{Bohn24} observes the central region of NGC 3256, a local merging LIRG, with JWST. This work detects outflowing warm \Htwo~near the nucleus of this system.
The emission of the outflowing region is fitted by a power-law index ranging from 4.5 to 5.9, consistent with shock-heating of the gas by the outflow and a warm-to-cold gas mass ratio of $4\%$, higher than MACS1931. \citet{Jones24} studies the nuclear starburst region of M83 using the JWST/MIRI instrument and finds a clumpy gas reservoir in the center of this spiral galaxy. The power-law index of this emission is $\rm n=5.75 \pm 0.64$, higher than this value for MACS1931 BCG, showing less excited \Htwo~in M83 center compared to MACS1931 BCG. The total \Htwo~traced warm gas mass in M83 is $\sim1\%$ of the CO traced cold gas mass, which is consistent with these values for MACS1931 BCG and the CGM tail. \citet{Jones24} speculates that star formation-induced shocks are responsible for the observed properties of this gas.

Galaxies within clusters and groups that are undergoing interactions with their companion galaxies or the IGM/ICM are affected by various physical processes that result in different excitation properties from isolated galaxies. Stephan’s Quintet is a nearby compact galaxy group with a large intergalactic gas filament that is mainly heated by shocks originating from the interaction with an intruder galaxy \citep{Appleton06}. The power-law index of the shock filament in Stephan’s Quintet ranges from 4.35 in the central parts of the shock to 4.85 in the outer regions of the filament, which are generally lower than the MACS1931 BCG power-law index. The warm gas mass fraction of the MACS1931 system is similar to the outer parts of the filament and less than the central parts of the filament, where the shock is supposed to be most intense \citep[$\sim15\%$;][]{Appleton17, Appleton23}.
\citet{Sivanandam14} reports Spitzer/IRS observations of warm \Htwo~tails in two cluster galaxies undergoing ram-pressure stripping with warm-to-cold gas mass ratio of $15\%-27\%$ (derived based on the reported warm-to-total gas mass ratio in the article). They conclude that the \Htwo~emission is mostly from shock-heated gas as a result of interactions with the ICM during ram-pressure stripping.

Given the complex nature of MACS1931 as a cool-core BCG hosting both an active AGN and a starburst, and the limited number of \Htwo~lines detected in our observations, identifying the dominant source of the \Htwo~emission remains challenging. It is likely that both starburst-driven processes and shocks contribute to the observed emission. However, fully constraining the role of shocks will require comprehensive modeling \citep{Appleton17} using additional \Htwo~lines and ionized gas tracers, to better characterize the shock properties and reveal their origin in this environment.

\subsection{Heating and cooling in MACS1931}

To understand the excitation of warm \Htwo~in MACS1931, several potential heating mechanisms must be considered. These include shocks, FUV radiation from massive stars, X-ray emission produced by AGN, and cosmic rays. Each of these mechanisms can contribute differently depending on the local gas conditions and timescales involved. With the available data, we can only estimate the X-ray contribution in \Htwo~heating in this paper. Furthermore, we explore the plausibility of the dissipation of the kinetic energy of warm \Htwo~resulting in the production of the cold gas traced with CO.

A possible source of \Htwo~heating is the X-ray emission from AGN \citep{Maloney96, Ogle10, Guillard12, Mackey19}. The BCG of MACS1931 has an AGN with an X-ray luminosity of $\sim 8 \times 10^{36}\, \mathrm{W}$ in the energy range of 0.7--8 keV \citep{Ehlert11}. Following the approach presented in \citet{Maloney96} and \citet{Ogle10}, we assume that 35\% of the absorbed X-ray flux in the energy range of 1--30 keV heats up the gas. With the standard AGN X-ray luminosity frequency dependence of $\nu^{-0.7}$ \citep{Lusso2016}, the ratio of the X-ray luminosity in the frequency range of 0.7--8 keV to the frequency range of 1--30 keV is $L(0.7-8\,\mathrm{keV}) / L(1-30\,\mathrm{keV}) = 1.83$. At a gas temperature of 515 K and density of $10^5 \, \mathrm{cm}^{-3}$ \citep{Ghodsi24}, the effective \Htwo~ionization parameter of the BCG is $\log \xi_{\mathrm{eff}} \sim -2.3$, resulting in the \Htwo~rotational cooling contributing $\sim 6\%$ to the total cooling \citep{Maloney96}. On the other hand, based on the extrapolation of the continuous temperature modeling in Section \ref{sec:Tex}, the luminosity ratio of the first three \Htwo~rotational lines to the total \Htwo~luminosity is $\rm L( H_2 S(1) - H_2 S(3)) / L(H_2) \sim 0.75$ \footnote{The predicted luminosities of the \Htwo S(2), \Htwo S(3), \Htwo S(4) lines based on the model are $1.2\times 10^{35}$ W, $3.1\times 10^{35}$ W, $7.6\times 10^{34}$ W, respectively.}. Considering all these factors, the maximum luminosity ratio of the $\rm H_2 S(1)-H_2S(3)$ lines to the X-ray luminosity in the energy range of 0.7--8 keV is 
$\rm L(H_2 S(1)-H_2S(3))/L_X(0.7-8\,\mathrm{keV}) = 0.03$. In the case of MACS1931, if we optimistically assume that all of the X-ray energy of the AGN is carried out to the location of the CGM tail, the upper limit of the ratio is $\sim 0.09$ and $\sim 0.04$ for the BCG and CGM, respectively. Hence, X-ray emission is not sufficient to explain the total \Htwo~emission.

\citet{Guillard09} suggests that the \Htwo~rotational lines originate from the dissipation of kinetic energy into molecular gas, which then cools down to the cold molecular gas temperature, contributing to the CO-emitting gas. The dissipation timescale of warm gas is calculated using the ratio of the kinetic energy to the \Htwo~rotational line luminosity:

\begin{equation}
    t_{\rm{diss}} = \frac{1}{2} \times M_{\rm{H_2}}^{\rm{warm}} \times \sigma ^2 \Big/ \sum_{\rm{j_{up}=3}}^{7} L_{\rm{S(j_{up})}}
\end{equation}

For MACS1931, we calculate the random kinetic energy of the warm gas using the masses reported in Section \ref{sec:Tex} and $\sigma^2 = 3 \times (\sigma^2_{\rm{S(5)}}-\rm spectral \, resolution^2)$ as the three-dimensional intrinsic velocity dispersion. The \Sfive~velocity dispersion is calculated from the FWHM reported in Table \ref{tab:MACS_lines}. The total luminosity of \Htwo~ is a summation over the observed and predicted luminosities of \Sone~to \Sfive~lines using the excitation model in Section \ref{sec:Tex}. The dissipation timescale is 0.26 Myr and 0.72 Myr for the MACS1931 BCG and CGM warm gas content, respectively. On the other hand, the dynamical timescale of gas ($\rm spatial \, resolution/\sigma$) in the BCG and CGM is 23.1 Myr and 10.9 Myr, respectively, calculated using the physical size of the resolution element divided by the velocity dispersion. Thus, the dissipation timescales are one order of magnitude shorter than the dynamical timescales, showing that it is plausible that the dissipation of the kinetic energy of the warm \Htwo~is the cooling source which leads to the formation of the CO-emitting gas. Stephan’s Quintet is an example of this scenario \citep{Guillard09}.

\subsection{The nature of the CGM tail structure}

The origin of the CGM tail of MACS1931 is not yet clear and different scenarios are suggested in the literature for the origin of this gas reservoir in MACS1931 BCG:

(\textit{i}) As discussed in \citet{Fogarty19} and supported by \citet{Ciocan21}, the kinematics and morphology of the cold molecular and ionized gas in the CGM reservoir suggest that the structure may have formed through the condensation of material uplifted by a past AGN outburst, now raining back onto the galaxy and fueling the current episode of star formation. Our JWST/MIRI observations show similar kinematics and spatial distribution between the \Htwo-traced warm and the CO-traced cold molecular gas components, indicating the same origin for these gas components.

(\textit{ii}) The X-ray study of MACS1931 presented in \citet{Ehlert11} and VLT/MUSE observations presented in \citet{Ciocan21} suggest that the gas content of the MACS1931 BCG and its CGM reservoir may originate from the condensation and cooling of the hot plasma in the intracluster medium. Our \Htwo~data do not support or oppose the ICM origin of the gas reservoir. Although the ionized emission lines detected in the JWST/MIRI data are not explored in this paper, we will jointly analyze the MUSE and full MRS data in the future to test this scenario. A metallicity consistent with that of the ICM would support the scenario of ICM condensation.

Both these interpretations require the infall of gas onto the BCG which may be explained by the precipitation model and the chaotic cold accretion model. In the precipitation model \citep{Voit17}, a local thermal instability in the environment of a galaxy (ICM or CGM) results in a sufficiently short cooling time, which make the hot gas to condense to a colder phase. This colder gas then precipitates onto the galaxy and the AGN activity creates a self-regulatory system to sustain this baryon cycle. In this model, a multiphase gas is expected in observations and the model allows filamentary precipitation. The chaotic cold accretion model \citep{Gaspari18} explains the baryon cycle focusing on kinematics and turbulence, where the hot halo is perturbed by turbulence and filaments of warm gas are condensed out of the hot halo. These filaments are then condensed into cold clouds that precipitate onto the galaxy via chaotic cold accretion. In this model, a multiphase gas with high velocity dispersion ($100-200\,  \rm{km\, s^{-1}}$) is expected. Measurements of the cooling time and free-fall time in addition to high resolution maps of the velocity dispersion of multiple emission lines are useful to distinguish between these scenarios. Our \Htwo~data shows that the warm molecular gas has similar spatial distribution and kinematics with the cold molecular gas and ionized gas, consistent with both models. Furthermore, our MIR \Htwo~data shows high velocity dispersion which is in line with the predictions of the chaotic cold accretion model. Looking ahead, we can use the upcoming high-resolution JWST MIR and NIR observations to study the clouds and filaments kinematics that provides critical insight into the nature of the accretion process.

(\textit{iii}) Chandra X-ray observations of this system show clear evidence of the oscillatory motion of the MACS1931 cluster core in the north-south direction as a result of a past cluster-scale merger, resulting in sloshing-induced gas stripping. \citet{Ehlert11} suggests that this gas reservoir may be the low entropy cool-core gas displaced through sloshing of the cluster center and affected by an AGN outburst later. The HST and Subaru Suprime-Cam image of MACS1931 BCG show a young stellar population in the south of the BCG without any $\rm H\alpha$ detected in this region that supports the scenario of the primary star-forming region moving northward \citep{Ehlert11}. The ALMA CO data and JWST \Htwo~data, similar to $\rm H\alpha$, do not show any significant emission in the south of the BCG which is consistent with the sloshing-induced gas stripping scenario.

(\textit{iv}) Another scenario suggests that the gas reservoir in the center of MACS1931 may originate from multiple minor mergers with satellite galaxies. However, this scenario is disfavored since satellite galaxies are expected to lose most of their gas as a result of ram pressure stripping while they move towards the cluster center \citep{Alberts22} and the amount of gas transferred in minor mergers is not enough to create such a massive gas reservoir around the BCG.

Although our \Htwo~analysis in this paper supports some of these interpretations, we cannot certainly rule out any of them. Future complementary JWST observations and analysis of ionized spectral lines would improve constraints on the thermal and dynamical properties of this system.

\citet{Castignani25} presents CO observations of three cool-core BCGs at $\rm z \sim 0.4$ (RX 1532, MACS 1447, and CHIPS 1911) that are among the most star-forming and gas-rich BCGs at this redshift, similar to MACS1931. They find extended cold molecular gas reservoirs for all of these BCGs and argue that these galaxies are fed by their molecular reservoirs, sustaining their high star formation rates. \citet{Castignani25} suggests different origins for these reservoirs. RX 1532 shows a tail-like reservoir with optical filaments, very similar to MACS1931. This reservoir is $\rm \sim 100 \, km \, s^{-1}$ redshifted compared to the galaxy, and the velocity dispersion is up to $\rm \sim 200 \, km \, s^{-1}$ higher within the BCG compared to the reservoir. They suggest that cooling flows from the ICM are the origin of this reservoir. The higher velocity dispersion of the MACS1931 CGM compared to the BCG itself might reflect a more complex environment than in the RX 1532 case. The CO morphology in MACS 1477 shows two peaks in the intensity map in addition to a tail-like reservoir, and a velocity discontinuity is visible between the gas reservoir and the BCG. The reservoir in MACS 1447 is suggested to originate from gas sloshing or ram-pressure stripping of gas in the cluster. MACS1931 shows a different CO morphology, thus, we cannot use similar arguments to \citet{Castignani25} to consider ram-pressure stripping or sloshing as the origin of the gas tail in MACS1931. The gas reservoir of CHIPS 1911 is suggested to be the result of the tidal disruptions of the galaxy because of a recent or ongoing merger event, supported by the two symmetric tails in optical, which is not seen in MACS1931. Overall, the origin of MACS1931's CGM tail might be similar to RX 1532, however, more in depth modeling is needed to confirm that.

\subsection{Caveats and future work}

It is noteworthy that the conversion factor $\alpha_{CO}$ depends on the metallicity and environment. Based on the VLT/MUSE observations of MACS1931 reported in \citet{Ciocan21}, the gas-phase metallicity of the CGM tail is slightly higher than the BCG, on average. Considering that \citet{Bolatto13} shows a clear anti-correlation between $\alpha_{CO}$ and metallicity, using a metallicity-dependent $\alpha_{CO}$ would be more appropriate. However, it is not possible due to the lack of a robust measurement of metallicity in the CGM tail. Thus, we might be overestimating the cold gas in the CGM tail, and our reported warm-to-cold mass ratios are thus lower limits on this value in the tail.

 It is important to note that the mass derived using CO lines is just a lower limit of the total cold molecular gas mass. As shown in \citet{Ghodsi24}, this system is CI-rich and parts of the molecular clouds are CO-poor, likely due to the high cosmic ray ionization rate. Hence, we consider the cold gas mass a lower limit and the warm-to-cold gas mass ratio an upper limit of the actual physical value.
 Furthermore, a more sophisticated excitation temperature modeling is needed for regions with non-thermal heating mechanisms. This modeling requires having more hydrogen rotational lines detected in the CGM tail and our accepted Cycle 4 JWST program will provide the observations needed to perform this modeling. Therefore, the LTE estimate just provides a conservative upper limit of the actual mass of the warm molecular hydrogen. Measuring the contribution of cosmic rays in gas heating requires having information about the polarization of the synchrotron component of the continuum emission. Future ALMA and VLA polarizations observations would be useful for this study.
 
 A comprehensive stellar population modeling and shock modeling is needed to understand the nature of the gas reservoir of this system. For this purpose, we will study the ionized lines and polycyclic aromatic hydrocarbons (PAHs) features observed with MUSE and JWST in future.
 Inferring a more universal conclusion about the nature of gas filaments in the center of cool-core galaxy clusters is possible through observing more similar objects to enlarge the sample to achieve a statistical significance of this finding.

%%%%%%%%%%%%%%%%%%%%%%%%%%%%%%%%%%%%%%%%%%%%%%%%%%%%%%%%%%%%%%%%%%%%%%%%%%%%%%%%%
\section{Conclusion}
\label{sec:Conclusion}

In this work, we study the temperature distribution of the molecular gas in the MACS1931-26 BCG and the CGM reservoir around it using JWST and ALMA. 
This represents one of the first studies to spatially resolve warm \Htwo~emission in the CGM at MIR wavelengths, providing new insight into the multiphase molecular gas beyond the central galaxy.
We use the Boltzmann distribution to model the excitation temperature of both BCG and CGM reservoir and a continuous temperature model for the BCG. A combined examination of the cold and warm phases of molecular gas leads us to the main conclusions:

\begin{itemize}
    \item 
    Cold and warm phases of gas share similar spatial distribution and kinematics in the BCG and the CGM tail up to the extent where CO is detectable. The CGM tail is redshifted by $\sim 300 \, \rm km \, s^{-1}$ in both CO and \Htwo~emission lines. This indicates that the warm and cold molecular gas components have the same origin.
    
    \item
    The \Htwo~lines FWHM is on average $\sim 120 \, \rm km \, s^{-1}$ broader than the CO lines in the BCG and CGM. The CGM shows higher line width than the BCG, on average, potentially pointing to a more turbulent medium in the CGM or the presence of multiple clouds with varying velocities in the line of sight.

    \item 
    We calculate the excitation temperature under the LTE assumption to be $515.6 \pm 0.8$ K in the BCG and $535.2\pm 1.9$ K in the CGM tail. The warm gas surface density is estimated to be $(6.3 \pm 0.3) \times 10^5 \, \rm M_{\odot}\,kpc^{-2}$ and $(2.0 \pm 0.1) \times 10^5 \, \rm M_{\odot}\,kpc^{-2}$ for the BCG and the CGM tail, respectively.

    \item 
    An optimistic estimate of the warm-to-cold gas mass ratio in the CGM reservoir ($1.9\%\pm 0.3\%$) is approximately equal to the BCG ($1.4\%\pm 0.2\%$), less than the SINGS galaxies, likely due to the CO-rich nature of MACS1931.

    \item
    X-ray heating is insufficient to fully explain the \Htwo~luminosity. Other heating mechanisms, such as shocks, are necessary. Our findings suggest that the dissipation of kinetic energy in the \Htwo-emitting gas plays a role in the formation of the CO-emitting gas.
\end{itemize}

Our analysis represents a step forward in understanding the role of the cold CGM in shaping galaxy evolution and the baryon cycle between the central galaxy and its surrounding environment. This paper is the first of a series of papers on the MACS1931 BCG. In the following work, we will discuss the ionized lines and the PAH features observed in the JWST MIR and NIR data.

%%%%%%%%%%%%%%%%%%%%%%%%%%%%%%%%%%%%%%%%%%%%%%%%%%%%%%%%%%%%%%%%%%%%%%%%%%%%%%%%%%%%%%%%%%%%%%%%%%%%%%%%%%

\begin{acknowledgments}
We acknowledge the support of the Canadian Space Agency (CSA) [23JWGO2A03]. LG, LK, and AWSM acknowledge the support of the Natural Sciences and Engineering Research Council of Canada (NSERC) through grant reference number RGPIN-2021-03046. AWSM acknowledges the support through the ESO Visitor Programme that facilitated the completion of this work. DD acknowledges support from the National Science Center (NCN) grant SONATA (UMO-2020/39/D/ST9/00720). DD acknowledges support from the Polish National Agency for Academic Exchange (Bekker grant BPN /BEK/2024/1/00029/DEC/1). TGB acknowledges support from the Leading Innovation and Entrepreneurship Team of Zhejiang Province of China (Grant No. 2023R01008). We acknowledges the support of Kirsten Larson,  David Law, and the rest of the STScI JWST MIRI team in the observational preparation and data reduction process. 
The authors thank Padelis Papadopoulos for scientific advice, and acknowledge helpful discussions with Javier Álvarez-Márquez, Julie Hlavacek-Larrondo, Megan Donahue, Jason Tumlinson, Bjorn Emonts, Vincenzo Mainieri, Darshan Kakkad, Luca Di Mascolo, and Thomas Lai.

\end{acknowledgments}

\vspace{5mm}
\facilities{This work is based on observations made with the NASA/ESA/CSA James Webb Space Telescope. The data were obtained from the Mikulski Archive for Space Telescopes at the Space Telescope Science Institute, which is operated by the Association of Universities for Research in Astronomy, Inc., under NASA contract NAS 5-03127 for JWST. These observations are associated with program \#3629. The JWST observations used in this study were retrieved from the Mikulski Archive for Space Telescopes (MAST) at the Space Telescope Science Institute. The analyzed dataset is available at \dataset[doi: 10.17909/8msq-td16]{https://doi.org/10.17909/8msq-td16}. We also use the following ALMA data: ADS/JAO.ALMA\#2016.1.00784.S and ADS/JAO.ALMA\#2017.1.01205.S. ALMA is a partnership of ESO, NSF, NINS, NRC, MOST, ASIAA, KASI, in cooperation with the Republic of Chile.}

\software{ We use the following software packages in this work: JWST science calibration pipeline \citep{2024zndo..10870758B}, JDAVis \citep{jdaviz2022}, \texttt{CASA 6.5.4-9} \citep{McMullin07}, CAFE spectral fitting code \citep{Diaz-Santos2025}, Astropy \citep{Astropy2018}, SpectralCube \citep{Ginsburg2019}, \texttt{emcee} \citep{Foreman13}, Matplotlib \citep{Matplotlib2007}, Numpy \citep{NumPy2011}, SciPy \citep{SciPy2020}, and \texttt{scikit-image} \citep{scikit-image}.}

\appendix
\section{JWST Data Calibration}
\label{sec:jwst_data_cal}

\subsection{JWST Calibration Pipeline}

We utilize the JWST calibration pipeline release 1.14.0 \citep{2024zndo..10870758B} in conjunction with the Calibration References Data System (CRDS) context \texttt{jwst\_1235.pmap} to process the MIRI MRS data. 
We followed the standard calibration procedure outlined in the MIRI MRS Batch Processing Notebook (D. Law \& K. Larson, 2024)\footnote{\href{https://github.com/STScI-MIRI/MRS-ExampleNB/blob/main/Flight\_Notebook1/MRS\_FlightNB1.ipynb}{https://github.com/STScI-MIRI/MRS-ExampleNB/blob/main/Flight\_Notebook1/MRS\_FlightNB1.ipynb}}. The standard pipeline method is divided into three components: the \texttt{Detector1Pipeline}, which performs basic detector-level corrections by converting raw ramp data into corrected countrate (slope) images; the \texttt{Spec2Pipeline}, which further calibrates these countrate products, converting them from DN/s to surface brightness (MJy\,sr$^{-1}$) for fully calibrated individual exposures; and the \texttt{Spec3Pipeline}, which combines data from multiple dithered exposures into a single 2D or 3D spectral product, along with a combined 1D spectrum.

For each step, we follow the default pipeline parameters with a few modifications. In stage 1, we set the jump rejection threshold to 100 to diminish the impact of bright objects and short ramps, and we enable the detection of large cosmic ray showers. In stage 2, we use the \texttt{mingrad} algorithm to run pixel replacement, extrapolating values for bad pixels to mitigate small 5–10\% negative dips in spectra of bright sources. We also opt for the standard pixel-wise background subtraction method, which directly subtracts the dedicated sky exposure from all six science exposures to remove background signals, an ideal method for faint, diffuse signals—such as the extended CGM tail in our case. For consistency, we also tested the alternative master background subtraction method available in Stage 3 of the pipeline, which instead infers and utilizes a background field within the science data. This approach yielded no significant improvement in the resulting SNR across the field of view. In stage 3, we adjust the outlier detection kernel size to $11\times1$ pixels for normalizing pixel differences and set the detection threshold for identifying bad pixels to 99.5\% of the normalized minimum pixel difference.

In addition to the MRS data, we calibrated the MIRI F2100W imaging data (used for astrometry correction - see \ref{ssec:astr_cor}) following the standard parameters outlined in the 31st JWebbinar calibration notebook provided by the Space Telescope Science Institute\footnote{\url{https://github.com/spacetelescope/jwebbinar_prep/blob/jwebbinar31/jwebbinar31/miri/MIRI_Imager_pipeline_demo-platform.ipynb}}. The imaging pipeline also consists of three stages: the \texttt{Detector1Pipeline}, identical to that of the spectral pipeline; \texttt{calwebb\_image2}, which applies further corrections and calibrations to produce fully calibrated exposures; and \texttt{calwebb\_image3}, which combines multiple exposures (e.g., dither or mosaic patterns) into a single rectified product. Modifications to the default parameters include reducing the rejection threshold to 5 in stage 1, and in stage 3, using the GAIA DR3 astrometric catalog \cite{2023A&A...674A...1G}, setting a minimum of 15 objects for matching, requiring at least 25 stars to align dithers, applying a Gaussian kernel for PSF centroiding, and doubling the default scaling factor applied to the derivative used to identify bad pixels.

\subsection{Post-Pipeline Stripe Removal}

In our MRS data, we observe dark vertical stripes in channels 1 and 2 when summing over all cube slices, which result in negative fluxes in extracted spectra of low SNR regions. These stripes, likely caused by Cosmic Ray (CR) showers, are present in the dedicated sky background and lead to such negative stripe bands after background subtraction in our Level 3 cubes. To address this, we follow the method outlined by \cite{2023Natur.618..708S}, first masking the position of our source (BCG+CGM tail) with a dilated mask (based on SNR a threshold of 2 in the H$_2$S(5) intensity map, see Section~\ref{sec:ferr_est}), then using \texttt{photutils} to create a 2D spatial stripe template extended along the spectral axis, and finally subtracting this template from the science cube to remove the artifacts.

\subsection{Astrometry Correction}
\label{ssec:astr_cor}

To accurately compare the JWST MIRI MRS data with ALMA observations, we must correct for astrometric discrepancies between the two observations. We start by comparing the source catolog derived from the calibrated adjacent MIRI image to the GAIA DR3 \citep{2023A&A...674A...1G} sources within the same field. We account for the proper motion of the GAIA sources to match the date of the JWST observations. We then use \texttt{Astropy}'s \texttt{match\_to\_catalog\_sky} function to match all sources in both catalogs with a maximum separation of 1$''$ from one another (2 object matches in total), and calculate a mean RA correction of 0.14$''$ and a declination correction of -0.01$''$. The uncertainties of these corrections, calculated as the RMS of the offsets between stars, are 0.01$''$ in RA and 0.03$''$ in declination. This is consistent with the MIRI pointing accuracy of about 14 mas \citep{Lallo2022}. This astrometric correction is then applied to the MRS data.

\section{Flux Uncertainty Estimations}
\label{sec:ferr_est}

As noted by \cite{2023AJ....166...45L}, the error cubes provided in Stage 3 of the JWST pipeline currently suffer from a bug in the uncertainty estimation process, resulting in values that are underestimated by a factor of 10 or more. To estimate flux errors, we proceed as follows. 
Starting from a calibrated Level 3 science cube, we mask out edge and source pixels—identified as those with SNR $>$ 2 in the H$_2$S(5) moment zero map relative to its error—to exclude emission from the BCG or CGM. We use H$_2$S(5) as it offers the clearest view of the CGM tail structure (see Figure~\ref{fig:mom1-2}).
Next, we apply a 6x6 pixel square dilation and a closing mask using functions from the \texttt{skimage.morphology} library (\citealt{scikit-image}).
Finally, we compute the root mean square (RMS) of the unmasked pixels in each wavelength slice and adopt it, in units of Jy$\,$pixel$^{-1}$, as the uncertainty for that given wavelength channel. To compute the error spectrum per aperture in Jy, we multiply the RMS per wavelength slice by the extraction aperture’s area in pixels. To account for systematic uncertainties—primarily arising from continuum subtraction—we sum the calculated error spectrum in quadrature with the RMS of line-free channels within each MRS channel. This method results in error bars that better align with expectations when analyzing spectral data.

%%%%%%%%%

\bibliography{bib}{}
\bibliographystyle{aasjournal}

\end{document}